# High-efficiency perovskite-organic blend light-emitting diodes featuring self-assembled monolayers as hole-injecting interlayers

*Murali Gedda,\*,† Despoina Gkeka,† Mohamad Insan Nugraha, Alberto D. Scaccabarozzi, Emre Yengel, Jafar I. Khan, Iain Hamilton, Yuanbao Lin, Marielle Deconinck, Yana Vaynzof, Frédéric Laquai, Donal D. C. Bradley and Thomas D. Anthopoulos\**

Dr. M. Gedda, Ms. D. Gkeka, Dr. M. I. Nugraha, Dr. A. D. Scaccabarozzi, Dr. E. Yengel, Dr. J. I. Khan, Prof. F. Laquai, and Prof. T. D. Anthopoulos

King Abdullah University of Science and Technology (KAUST),

KAUST Solar Center (KSC), Thuwal 23955-6900, Saudi Arabia

†These authors have contributed equally.

E-mail: murali.gedda@kaust.edu.sa; thomas.anthopoulos@kaust.edu.sa;

M. Deconinck, and Prof. Y. Vaynzof

Dresden Integrated Center for Applied Physics and Photonic Materials (IAPP) and Center for Advancing Electronics Dresden (cfaed), Technische Universität Dresden, 01062 Dresden, Germany

Dr. M. I. Nugraha

Research Center for Advanced Materials, National Research and Innovation Agency (BRIN), South Tangerang, Banten 15314, Indonesia

Dr. I. Hamilton and Prof. D. D. C. Bradley

King Abdullah University of Science and Technology (KAUST),

Division of Physical Science and Engineering, Thuwal 23955-6900, Saudi Arabia

The high photoluminescence efficiency, color purity, extended gamut, and solution processability make low-dimensional hybrid perovskites attractive for light-emitting diode (PeLED) applications. However, controlling the microstructure of these materials to improve the device performance remains challenging. Here, the development of highly efficient green PeLEDs based on blends of the quasi-2D (q2D) perovskite, $PEA_2Cs_4Pb_5Br_{16}$, and the wide bandgap organic semiconductor 2,7 dioctyl[1] benzothieno[3,2-b]benzothiophene ($C_8$-BTBT) is reported. The presence of $C_8$-BTBT enables the formation of single-crystal-like q2D $PEA_2Cs_4Pb_5Br_{16}$ domains that are uniform and highly luminescent. Combining the PEA$_2$Cs$_4$Pb$_5$Br$_{16}$:C$_8$-BTBT with self-assembled monolayers (SAMs) as hole-injecting layers (HILs), yields green PeLEDs with greatly enhanced performance characteristics, including external quantum efficiency up to 18.6%, current efficiency up to 46.3 cd A$^{-1}$, the luminance of 45 276 cd m$^{-2}$, and improved operational stability compared to neat PeLEDs. The enhanced performance originates from multiple synergistic effects, including enhanced hole-injection enabled by the SAM HILs, the single crystal-like quality of the perovskite phase, and the reduced concentration of electronic defects. This work highlights perovskite:organic blends as promising systems for use in LEDs, while the use of SAM HILs creates new opportunities toward simpler and more stable PeLEDs.

1. Introduction

Metal halide perovskites have attracted intense scientific interest in recent years for application in the broader field of optoelectronics, including solar cells,[1-3] light-emitting diodes (LEDs),[4] lasers,[5] and photodetectors.[6] The combination of outstanding optical and electrical characteristics, including high photoluminescence quantum yield (PLQY), tunable bandgap, high charge carrier mobility, and high color purity, makes metal halide perovskites particularly attractive for use in perovskite-based LEDs (PeLEDs) for various applications in optical displays and lighting.[7-9] Indeed, progress in this area has been dramatic, with the external quantum efficiency (EQE) of PeLEDs increasing rapidly from 0.1% to ≈28% in recent years.[10-15] These strides are attributed primarily to improvements in trap passivation via compositional engineering, balanced carrier injection, an increase in the radiative carrier recombination, and improvements in light out-coupling.[9, 10]

Among the various materials studied, quasi-2D (q2D) hybrid perovskites, also known as Ruddlesden–Popper (RP) perovskites, attract increasing attention due to their enhanced environmental stability[16, 17] and high PLQY.[18, 19] The majority of hybrid RP perovskites reported to date consist of a mixture of 2D domains characterized by different layer numbers (*n*-value) due to the low formation energies.[20] In such complex solids, the 2D domains exhibit quantum-well (QW) like characteristics with a high exciton binding energy,[21, 22] with efficient energy transfer between domains of different *n*-value. It has recently been shown that energy can funnel from the high-bandgap domains (low *n*-value) to the lower bandgap domains (higher *n*-value), ultimately leading to efficient bimolecular recombination and higher PLQY.[22-24] However, when it comes to applications in PeLEDs, both the microstructure and morphology of the perovskite layer, as well as its interactions with the charge injection layers,[4, 9, 10, 25-27] tend to govern the overall efficiency of the device. For example, the presence of structural defects can result in nonemissive carrier recombination during electrical excitation, while the existence of large injection barriers can lead to unbalanced carrier transport, reduced recombination, and lower EQE.[28-30] Thus, the careful engineering of all device components (i.e., light-emitting layer, interlayers and photonic structure) is necessary to improve the overall performance of PeLEDs.

Introducing molecular additives into the perovskite inks has recently emerged as an effective strategy toward improving the quality of the perovskite layer and the overall performance of perovskite devices.[31-33] Control over crystallization and growth during solution processing of the perovskite layer, defect passivation, and suppressed ion migration, are some of the important functions that selected additives offer. Organic ligands, ammonium halides with large organic cations, or polymer matrices are examples of materials used to enhance the radiative recombination rate within the perovskite layer by improving the spatial confinement of electrons and holes.[34-36] Therefore, identifying molecules/additives that facilitate adequate control over the nucleation and growth of the perovskite layer while simultaneously suppressing the concentration of electronic defects could lead to further improvements in the PeLED performance. For example, Zhao et al. used an insulating polymer blended with the q2D perovskite $(NMA)_2(FA)Pb_2I_7$ to produce red-emitting PeLEDs with a maximum EQE of 20.1%.[13] Adopting a similar approach but using an organic conjugated small-molecule (SM) instead, we demonstrated controlled crystallization of the q2D perovskite $(PEA)_2PbBr_4$ and its application in nonvolatile memory devices,[37] further highlighting the potential of blends for improving the device operation. In parallel to these efforts, work toward charge transporting interlayers has also been advancing. One recent intriguing development involves the use of engineered SAMs as hole-transporting materials in perovskite and organic photovoltaics.[38-42] Despite the tremendous promise, such SAMs have yet to be deployed in PeLEDs.

Inspired by these developments, we set out to explore the use of 2,7 dioctyl[1] benzothieno[3,2-b]benzothiophene ($C_8$-BTBT) as a molecular additive for the light-emitting perovskite layer in $PEA_2Cs_4Pb_5Br_{16}$-based LEDs. The motivation for choosing $C_8$-BTBT is its high solubility, good environmental stability, and highly crystalline-layered thin-film morphology that resembles those found in q2D halide perovskites. We have previously[37] shown that $C_8$-BTBT affects the crystallization and growth kinetics of the $PEA_2Cs_4Pb_5Br_{16}$ phase during solution processing leading to the formation of larger crystalline domains and fewer grain boundaries. Layers of $PEA_2Cs_4Pb_5Br_{16}$:$C_8$-BTBT blends processed from optimized formulations exhibit reduced nonradiative recombination and a longer exciton decay lifetime. PeLEDs made using $PEA_2Cs_4Pb_5Br_{16}$:$C_8$-BTBT and poly(3,4-ethylenedioxythiophene) polystyrene sulfonate (PEDOT:PSS), as the emissive and hole-injecting layer (HIL), respectively, exhibit green emission with a maximum EQE of 3.32%, a peak brightness of 7049 cd m$^{-2}$, and maximum current efficiency of 7.74 cd A$^{-1}$. Blend PeLEDs exhibit enhanced operational stability (15%) compared to neat perovskite diodes. Replacing the conventional PEDOT:PSS with engineered SAM-based hole-injecting layers (HILs) yields PeLED with drastically increased EQE of 18.6%, a peak brightness of 45 276 cd m$^{-2}$, and maximum current efficiency of 46 cd A$^{-1}$. These performance characteristics are comparable to the best PeLED data reported to date (Table S1, Supporting Information). Our work highlights the promise of perovskite:organic small-molecule blends for application in stable and highly efficient PeLEDs while demonstrating the enormous potential of SAMs as carrier-injecting interlayer materials for PeLEDs.

## 2 Results and Discussion

Solution-processed layers of the q2D perovskite PEA$_2$Cs$_4$Pb$_5$Br$_{16}$ are known to consist of a mixture of phases with different dimensionality that contains PbBr$_6$ layers sandwiched between phenetylammonium (PEA) layers (**Figure** 1a).[43, 44] For our experiments, the PEA$_2$Cs$_4$Pb$_5$Br$_{16}$ precursors and C$_8$-BTBT solutions were prepared separately and mixed in a 75:25 perovskite:small-molecule volume ratio (vol%) to form the blend formulation. The particular vol% ratio was chosen because it was found to yield films with the highest uniformity and coverage. The substrates were cleaned with utmost care to ensure smooth and contamination-free ITO surfaces. Details of the cleaning procedure are described in Experimental Section. The atomic force microscopy (AFM) scanned topography of ITO (Figure S1a, Supporting Information) and the RMS roughness (1.52 nm) measured from the height histogram profile in Figure S1b, Supporting Information confirms the clean (i.e., absence of visible contaminants) and smooth ITO surface. Further, as-prepared blend solutions were then spin-coated at room temperature inside a nitrogen-filled glovebox on clean glass substrates, followed by thermal annealing. Details of the solution preparation and spin coating are provided in Experimental Section. We note that, unlike our previous work, we employed toluene as an antisolvent during the fabrication of the blended PEA$_2$Cs$_4$Pb$_5$Br$_{16}$:C$_8$-BTBT samples. This significantly impacts the vertical compositional profiles of the deposited films. Specifically, due to the high solubility of C$_8$-BTBT in toluene, the previously observed vertical stratification of C$_8$-BTBT to the surface of the PEA$_2$Cs$_4$Pb$_5$Br$_{16}$:C$_8$-BTBT blends is avoided. To confirm this, we performed X-ray photoemission spectroscopy (XPS) depth profiling experiments on samples fabricated with and without an antisolvent (Figure S2, Supporting

Information). In the blend film, fabricated without the use of an antisolvent (Figure S2a, Supporting Information), the surface consists primarily of C$_8$-BTBT, evidenced by the high atomic percentages of carbon (≈62%) and sulfur (≈3%). On the other hand, when using toluene as an antisolvent, the carbon signal at the surface originates from adventitious carbon that is removed with the first etching step, resulting in only a 12% of carbon contribution throughout the film (Figure S2b, Supporting Information). The sulfur content in the antisolvent fabricated films is ≈0.5% both at the surface and at the bulk of the blend layers, similar to the neat perovskite system (Figure S2c, Supporting Information). These findings reveal the effective removal of molecular additives from the perovskite surface using antisolvent.

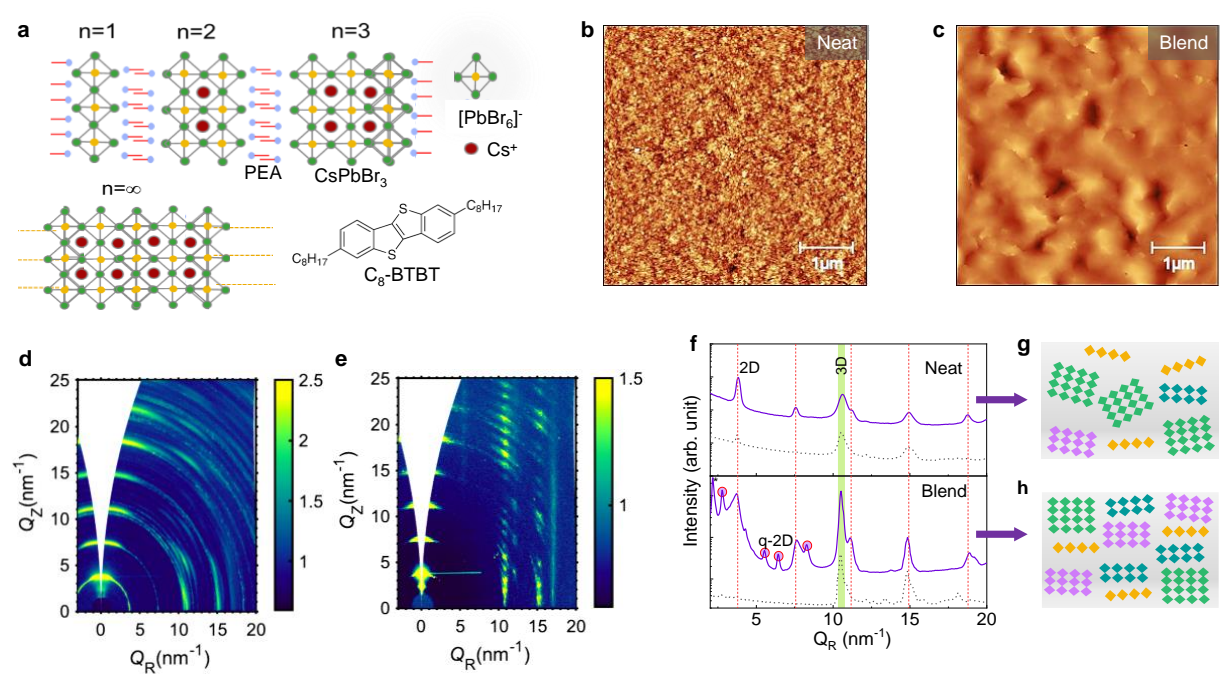

Figure 1. a) Schematic representation of quasi-2D perovskite with different layers number, n, and molecular structure of organic small-molecular additive C$_8$-BTBT. AFM topography and GIWAXS diffractograms of b, d) neat and c, e) blend quasi-2D perovskites. f) Out of plane (solid line) and in-plane (dash line) diffraction patterns of neat (top frame) and blend (bottom frame) films. Graphical representation of crystallite orientation quasi-2D perovskite films of g) neat and h) blend.

The topography of the resulting films was investigated using AFM (Figure 1b,c). Neat PEA$_2$Cs$_4$Pb$_5$Br$_{16}$ films appear polycrystalline with a smaller grain-like morphology and are characterized by a high density of grain boundaries. On the other hand, PEA$_2$Cs$_4$Pb$_5$Br$_{16}$:C$_8$-BTBT blends are significantly more dense and homogeneous and feature larger crystalline domains with fewer grain boundaries (Figure S3a,b, Supporting Information). Further analysis of the AFM data (Figure S3c,d, Supporting Information) reveals the formation of larger domains (≥1 µm in length and ≈37 nm in height), further highlighting the drastic impact of $C_8$-BTBT on the morphology of the resulting perovskite layer.

The crystal microstructure and texture of the neat $PEA_2Cs_4Pb_5Br_{16}$ and $PEA_2Cs_4Pb_5Br_{16}$:$C_8$-BTBT blend layers were further analyzed using grazing-incidence wide-angle X-ray scattering (GIWAXS) measurements. As shown in Figure 1d,e, the data obtained at an incident angle of 1° reveal major structural differences between the two samples. The neat $PEA_2Cs_4Pb_5Br_{16}$ layer shows long arcs of intensity in the GIWAXS pattern, indicating that the crystallites are isotropically distributed without any preferred orientation (Figure 1d).[45] In contrast, the $PEA_2Cs_4Pb_5Br_{16}$:$C_8$-BTBT blend (Figure 1e) shows sharp peaks with small arc lengths along the $Q_z$-axis, indicating the presence of a strongly textured and preferentially oriented film. Due to the large spacing induced by the PEA molecules located between the $PbBr_6$ planes in the perovskite, the 2D phases appear at $Q$ values (scattering momentum transfer) below 10 nm$^{-1}$. Figure 1f shows the in-plane (dashed line) and out-of-plane (solid line) diffraction patterns of neat and blend films. For clarity, the vertical red dashed lines indicate the diffraction peaks attributed to the pure 2D phase of PEA2PbBr4 with $n = 1$. In the case of the neat $PEA_2Cs_4Pb_5Br_{16}$, shown in Figure 1f, only one diffraction peak can be correlated with the 3D perovskite phase. As seen in Figure 1d, the arc from the diffraction is almost isotropic, indicating that the lead-bromide sheets in the q2D phase are oriented randomly with a slight perpendicular (edge-on) orientation concerning the substrate (Figure 1g).

The $PEA_2Cs_4Pb_5Br_{16}$:$C_8$-BTBT blend (Figure 1f, bottom frame), on the other hand, shows several scattering peaks that correspond to the q2D phase (marked with purple dots). The existence of sharp profiles and short arc length indicate that the $PbBr_6$ sheets are oriented parallel to the substrate, as schematically depicted in Figure 1h. The presence of large and dense domains comprised of q2D perovskite phases with varying n values is known to promote efficient energy transfer from the high bandgap phases (i.e., $n = 1, 2…$) to lower bandgap phases (i.e., bulk/3D domains) resulting to higher radiative recombination that makes them attractive for light-emitting applications.[20, 22, 23] In an effort to visualize the spatial distribution of the various phases, hyperspectral PL measurements were carried out on neat $PEA_2Cs_4Pb_5Br_{16}$ and $PEA_2Cs_4Pb_5Br_{16}$:$C_8$-BTBT blend films. Figure S4a,b, Supporting Information shows 2D PL maps measured for the two samples, together with discrete PL spectra extracted from the different locations highlighted on each map (Figure S4c,d, Supporting Information). Blending has a little perceptible effect on the PL spectrum, which is ascribed entirely to emission from PEA$_2$Cs$_4$Pb$_5$Br$_{16}$. Light emission from the blend film appears more uniform than the neat layer, as evident from the optical photographs (Figure S4e,f, Supporting Information) taken under excitation with UV light, further highlighting the positive influence of C$_8$-BTBT.

To investigate the electroluminescence (EL) properties of the neat PEA$_2$Cs$_4$Pb$_5$Br$_{16}$ and the PEA$_2$Cs$_4$Pb$_5$Br$_{16}$:C$_8$-BTBT blend films, we fabricated multilayer bottom-emitting PeLEDs comprised of ITO/PEDOT:PSS/perovskite(neat/blend) layer/ 2,2′,2″-(1,3,5-Benzinetriyl)-tris(1-phenyl-1-H-benzimidazole)(TPBi)/Ca/Al (**Figure** 2a, inset). Figure 2a,b shows representative sets of the current density−luminance−voltage ($J-L-V$) characteristics and the EL spectra as a function of the voltage of both devices, respectively. Table S2, Supporting Information summarizes the operating parameters of both types of PeLEDs. Interestingly, blend PeLEDs exhibit reduced hysteresis between forward and return sweeps (Figure S5a, Supporting Information). Mobile ions, ferroelectric dipoles and charge trapping at interface/grain boundaries are a few mechanisms from which the $J$–$V$ hysteresis has been proposed to originate.[46, 47] However, the clockwise hysteresis of our PeLED $J$–$V$ curves (arrows in Figure S5a, Supporting Information show the direction of the hysteresis) suggest the dominant role of charge traps, most likely present at the interface of the grain boundaries. The EL emission spectra of both devices are almost identical, with the $\lambda_{max}$ centered at ≈515 nm with a full width at half maximum (FWHM) of ≈25 nm (Figure S5b, Supporting Information). Importantly, the emission spectra of both PeLEDs remain unchanged with increasing bias voltage (Figure 2b), as are their Commission Internationale de l'Eclairage (CIE-1931) color coordinates (Figure 2c), yielding (0.12, 0.73) and (0.11, 0.74) for neat and blend PeLEDs, respectively. The high purity of the green color emitted by these PeLEDs makes them highly attractive for applications in optical displays as it enables access to a significantly wider color gamut.

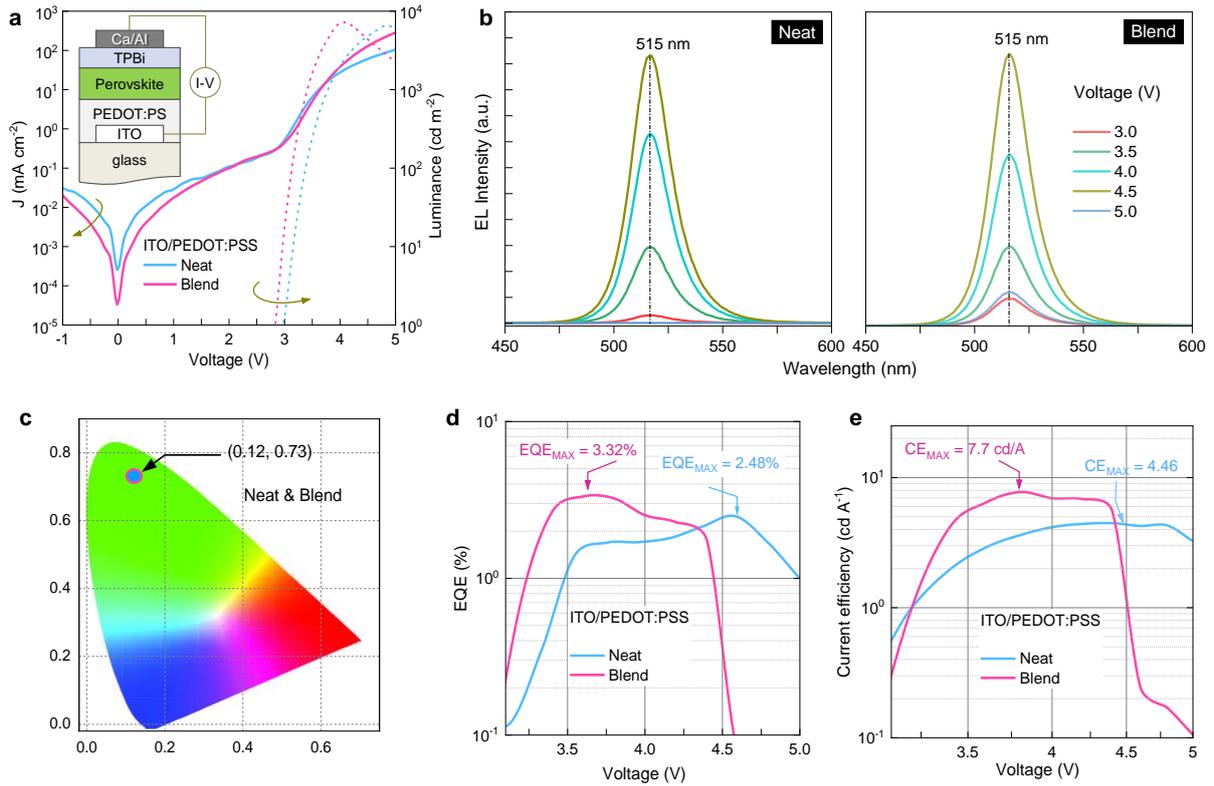

Figure 2. Operating characteristics of PeLEDs based on neat PEA$_2$Cs$_4$Pb$_5$Br$_{16}$ and PEA$_2$Cs$_4$Pb$_5$Br$_{16}$:C$_8$-BTBT blend as the emissive layer and PEDOT:PSS as the hole injecting interlayer. Data are shown for: a) Current density – luminance – voltage; b) Voltage-dependent electroluminescence spectra; c) Corresponding CIE(x, y) 1931 chromaticity coordinates; d) External quantum efficiency (EQE) versus voltage and e) Current efficiency versus voltage.

The neat PEA$_2$Cs$_4$Pb$_5$Br$_{16}$ and PEA$_2$Cs$_4$Pb$_5$Br$_{16}$:C$_8$-BTBT blend based PeLEDs exhibit similar turn-on voltages ($V_{on}$) of ≈2.75 V (Figure 2a) although their overall performance differs significantly. Specifically, the maximum luminance (cd m$^{-2}$) (Figure 2a), EQE (Figure 2d), and current efficiency (cd A$^{-1}$) (Figure 2e) values of the neat PEA$_2$Cs$_4$Pb$_5$Br$_{16}$ PeLED are 6485 cd m$^{-2}$, 2.5%, and 4.5 cd A$^{-1}$, respectively. Replacing the neat PEA$_2$Cs$_4$Pb$_5$Br$_{16}$ with the PEA$_2$Cs$_4$Pb$_5$Br$_{16}$:C$_8$-BTBT blend yields PeLEDs with significantly improved performance, exhibiting a maximum luminance of 7049 cd m$^{-2}$ (Figure 2a), maximum EQE of 3.32% (Figure 2d), a peak current efficiency of 7.74 cd A$^{-1}$ (Figure 2e) and higher EL intensity (Figure S5b, Supporting Information). The enhanced performance is attributed to the improved quality of the perovskite phase due to the formation of larger crystalline domains and the presence of fewer grain boundaries (Figure 1b) that result in reduced operating hysteresis (Figure S5a, Supporting Information). To this end, the presence of grain boundaries in hybrid perovskites has indeed been linked to charge trapping, and hysteretic behavior[48] and hence could explain the larger hysteresis seen in neat PEA$_2$Cs$_4$Pb$_5$Br$_{16}$ devices (Figure S5a, Supporting Information). Based on these results, we conclude that the addition of C$_8$-BTBT into the PEA$_2$Cs$_4$Pb$_5$Br$_{16}$ enables the growth of perovskite layers with improved microstructural characteristics, ultimately leading to better-performing PeLEDs.

It has been recently shown that enhancing the hole injection in PeLEDs can also lead to improvements in performance[29] by balancing carrier injection and recombination, ultimately leading to enhanced radiative recombination. In an attempt to improve the hole injection in our best performing PEA$_2$Cs$_4$Pb$_5$Br$_{16}$:C$_8$-BTBT blend PeLEDs, we replaced the PEDOT:PSS layer (see inset in Figure 2a) with two high ionization potential SAMs, namely 2PACz and Br-2PACz (see Experimental Section for chemical names) shown in **Figure** 3a, as molecularly-thin HILs. Notably, both SAMs can be functionalized directly onto ITO (Figure 3a) from solution and have recently been utilized as hole-extracting interlayers in state-of-the-art organic photovoltaics (OPVs).[39, 40] The energy levels of each device component, including electrodes and active layers, are presented in Figure 3b. The energy levels of the PEA$_2$Cs$_4$Pb$_5$Br$_{16}$ perovskite were obtained via photoelectron spectroscopy in air (PESA) measurements combined with the optical bandgap extracted from the absorption spectra (Figure S6, Supporting Information). The lower-lying work function (6 eV) of the Br-2PACz functionalized ITO electrode (ITO/Br-2PACz) is expected to facilitate improved hole injection as compared to bare ITO (4.9 eV) and ITO/2PACz (5.5 eV) anode electrodes. Figure 3c shows the EL spectra measured for the PEA$_2$Cs$_4$Pb$_5$Br$_{16}$:C$_8$-BTBT-based PeLEDs utilizing the two SAMs. Under 3.5 V bias, the PeLEDs emit bright green light with the λ$_{max}$ remaining at ≈515 nm. Devices with Br-2PACz show higher EL intensity (inset photograph in Figure 3c) with a narrower FWHM (≈23 nm) as compared to 2PACz-based PeLEDs (≈25 nm). The EL spectra of both devices remain unchanged for bias in the range of 3 to 5 V, highlighting the stable green light-emitting character of the PeLEDs (Figure S7, Supporting Information). The CIE color coordinates calculated from the EL spectra of 2PACz and Br-2PACz-based blend PeLEDs yielded (0.12, 0.71) and (0.10, 0.74), respectively, highlighting the high purity of the emitted green light.

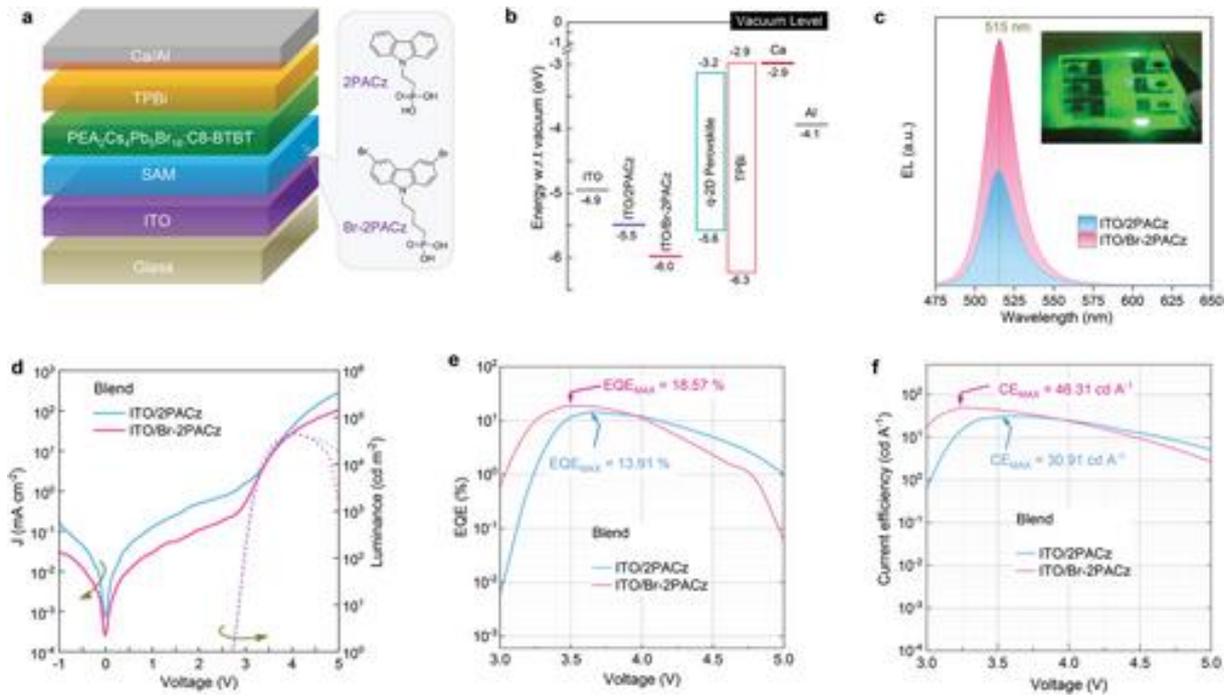

Figure 3. PeLED device architecture and key performance characteristics for blend based devices with 2PACz and Br-2PACz SAM hole injection layers: a) PeLED schematic and chemical structures of the hole-injection layers; b) Energy level diagram for the device functional layers; c) Electroluminescence spectra with inset photograph of the best-performing Br-2PACz/blend device; d) current density – luminance – voltage data; e) external quantum efficiency as a function of applied bias; and f) Current efficiency–voltage characteristics.

The *J–L–V* characteristics and corresponding EQE-V and CE-V plots of the blend PeLEDs with SAMs as HILs are presented in Figure 3d–f. Blend PeLEDs based on 2PACz and Br-2PACz HILs exhibit similar turn-on voltage around 2.6 V (determined as the driving voltage at ≈1 cd m$^{-2}$ luminance), a maximum luminance of 42 746 and 45 276 cd m$^{-2}$, the maximum current efficiency of ≈30.9 and 46.3 cd A$^{-1}$ (at a bias of ≈3.5 V) and maximum EQE of ≈13.9 and 18.6%, respectively. Within 0.5 V from the PeLED turn-on, a sharp rise in luminance from 1 to 10$^4$ cd m$^{-2}$ is attained. These results suggest that adding C$_8$-BTBT in combination with SAMs in the PeLED modulates the carrier injection which in turn increases bimolecular recombination, ultimately resulting in improved PeLED performance. On the other hand, the lower current density measured for ITO/Br-2PACz-based PeLEDs, together with the higher EQE and current efficiency (Figure 3d–f), suggests a more balanced carrier injection, which may be responsible for the improved bimolecular recombination and higher overall performance. In comparison, neat PeLEDs with 2PACz and Br-2PACz SAMs HILs show moderate operating characteristics with a maximum luminance of 39 116 and 45 078 cd m$^{-2}$, a maximum current efficiency of ≈18.3 and ≈26.8 cd A$^{-1}$, and maximum EQE values of ≈9.3 and ≈14.1%, respectively (Figure S8a–d, Supporting Information). This is an important observation and highlights the key role of C$_8$BTBT in improving the morphology and microstructure of the perovskite layer with a positive impact on the PeLED performance. Despite the lower performance of neat PeLEDs utilizing SAMs, the devices exhibit superior performance than neat PeLEDs made with PEDOT:PSS (Figure S9, Supporting Information). Moreover, SAM-based PeLEDs exhibit reduced hysteresis in their *J–V* characteristics, an effect particularly pronounced in Br-2PACz-based blend PeLEDs where the hysteresis is almost negligible (Figure S8e,f, Supporting Information). These findings conclude that replacing PEDOT:PSS with SAM HILs in blend PeLEDs leads to fewer interface traps, possibly due to reduced grain boundaries, hysteresis-free operation, higher luminance and increased EQE. This hypothesis is supported by experimental data presented so far on the improvement of the blend microstructure and includes AFM (Figure 1b,c and Figure S3, Supporting Information) and GIWAXS (Figure 1d–g).

In an effort to quantify the influence of the SAM HILs on the morphology of the neat PEA$_2$Cs$_4$Pb$_5$Br$_{16}$ and blend PEA$_2$Cs$_4$Pb$_5$Br$_{16}$:C$_8$-BTBT layers, we performed additional AFM measurements (Figure S10, Supporting Information). Blend layers appear slightly rougher than neat ones and consist of larger crystalline domains (Figure S10a,b, Supporting Information). Clear differences in the surface topography of the blend PEA$_2$Cs$_4$Pb$_5$Br$_{16}$:C$_8$-BTBT layers were observed for the three HILs studied. This, however, is not the case for the neat PEA$_2$Cs$_4$Pb$_5$Br$_{16}$ layer, with its topography remaining almost unaltered for all three HILs (i.e., ITO/PEDOT:PSS, ITO/2PACz, ITO-Br-2PACz) (Figure S10a, Supporting Information). We conclude from the AFM images and associated height histograms (Figure S10c,d, Supporting Information) that the blend films deposited on ITO/2PACz and ITO/Br-2PACz form larger size domains than the neat perovskite, with the average height distributions centered at ≈127 and ≈132 nm, respectively.

We performed scanning electron microscopy (SEM) measurements to gain deeper insights into the surface morphology of the various layers. As shown in Figure S11, Supporting Information, all the neat films display similar morphology with a key feature the significantly smaller crystalline grains. Specifically, neat films are characterized by much smaller grains the dimensions of which are difficult to resolve even at higher magnifications. Well-resolved grain morphology of neat films is obtained via the AFM measurements shown in Figure S3a,

Supporting Information. On the other hand, blend layers feature consistently larger grains (Figure S11, Supporting Information). Higher magnification scans of blend layers reveal that each grain is polycrystalline and consists of stacks of crystalline domains. This intriguing feature most likely underpins the enhanced efficiency measured for blend PeLEDs, as the larger crystalline domains most likely contain fewer grain boundaries and as such fewer trap states.

To understand the origin of these morphological differences, we studied the surface energy of each ITO/HIL system using water contact angle measurements (Figure S12a–c, Supporting Information). Using the Owens, Wendt, Rabel, and Kaelble (OWRK) equation,[49, 50] the surface energies for ITO/2PACz, ITO/Br-2PACz, and ITO/PEDOT:PSS were calculated (Figure S12d, Supporting Information). The ITO/2PACz and ITO/Br-2PACz electrodes exhibit similar surface energies, yielding 36.8 and 38.1 mN m$^{-1}$. The ITO/PEDOT:PSS electrode, on the other hand, appears significantly more hydrophilic leading to much higher surface energy (114.5 mN m$^{-1}$), which explains the different wetting and drying dynamics of the perovskite layers with the different HILs (Figure S10d, Supporting Information). Since the atomic compactness, crystal domain size, and morphology of the ensuing perovskite layer is known to depend strongly on the surface energetics of the underling material(s) system, we argue that the proposed SAM HILs provide a unique method to tune the surface energy with atomic precision via simple chemical substitutions on the periphery of the SAM molecule. Such atomic substitutions can in principle be designed to enable enhanced interactions with the under-coordinated lead ions in the perovskite.[51] Thus, SAM-based HILs could enable both precise control over the surface energetics, as well as the electronic properties, of the ITO electrode interface that is often inaccessible through the use of conventional thin-film HIL materials. Overall, the use of SAM HILs can be seen as a promising technology in terms of device performance, manufacturability and eco-friendliness due to its optimal material utilization.

To understand the influence of the various HILs on the photo-excited state dynamics within the PEA$_2$Cs$_4$Pb$_5$Br$_{16}$:C$_8$-BTBT layers, we performed further optical spectroscopy studies. The linear absorption spectra of neat and blend films grown on ITO/SAM substrates (Figure S13a, Supporting Information) show two dominant exciton absorption peaks located at 403 and 432 nm, confirming the primary film components to be PbBr$_6$-based single and bilayer ($n = 1$, 2) quantum wells (QWs) of PEA$_2$Cs$_4$Pb$_5$Br$_{16}$.[43, 52] The narrow exciton absorption peaks also suggest that the electronic coupling between the stacked QWs is relatively weak. Another small absorption peak at around 462 nm indicates the presence of a low-concentration of $n = 3$ QWs, while the absorption at 513 nm is associated with the presence of a bulk phase (i.e., $n \geq$ 5).[52] This self-doping and carrier localization effect can be clearly observed in transient absorption (TA) spectra, where photo-induced changes in the absorbance ($\Delta A$) spectrum of the q2D film are probed upon excited with a femtosecond laser pulse. The acquired spectra at different time delays of both the neat and blend samples are presented in **Figure** 4a,b. We observe energy transfer from $n = 1$ phase (400–410 nm) to $n \geq 5$ phase (504–524 nm) as indicated by the respective photobleaching (PBs) features corresponding to excitonic populations. This is supported by the observation of an initial decay of the $n = 1$ PB accompanied by a fast rise of the $n \geq 5$ PB. Interestingly, the blend sample did not exhibit such an energy transfer signature when monitoring the spectral evolution at $n \geq 5$ region. In our recently published study,[53] we have indeed observed a strong photo-induced absorption signal but only for the $n = 1$ layered perovskite. In the present q2D perovskite system, the ground state bleaching dominates and exhibits mostly energy transfer followed by exciton recombination. A closer look at the dynamics tracked at $n = 1$ and $n \geq 5$ PBs (Figure 4c and Figure S13b,c, Supporting Information) indicates that the energy transfer time is around 1–2 ps. We observed distinct dynamics when comparing the dynamics of neat and blend films across the $n \geq 5$ spectral region (Figure S13b,c, Supporting Information). More precisely, the blend film exhibits a faster exciton decay compared to the neat film, suggesting improved crystallinity of the former. On longer time scales, here >100 ps, exciton recombination dominates the decay dynamics and is found to be identical for both neat and blend samples. Moreover, the PL spectra for both neat and blend films show a single narrow peak at 515 nm, confirming the presence of fast and efficient energy transfer from low-$n$ to high-$n$ phases (Figure 4d and Figure S13d, Supporting Information).

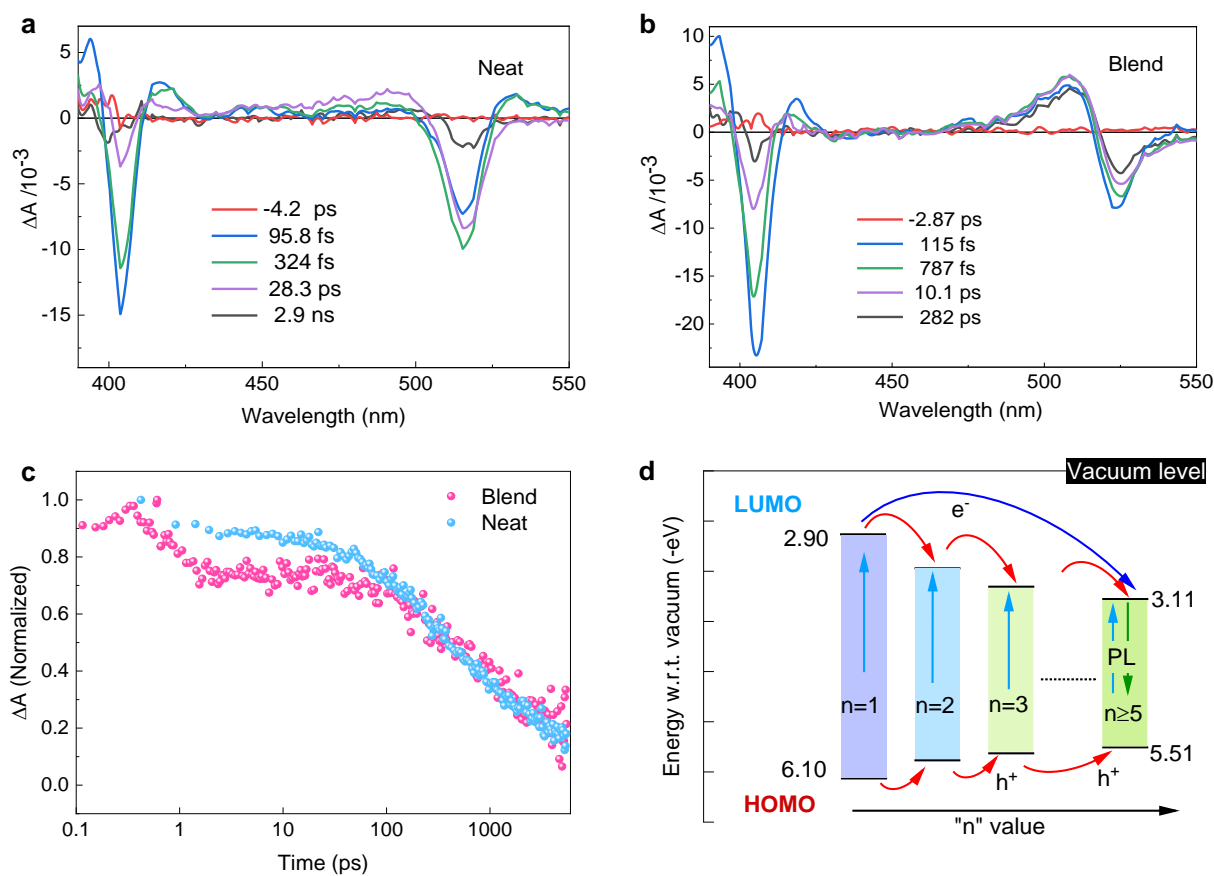

Figure 4. Photo-induced absorption spectra (ΔA) and decay kinetics for neat and blend perovskite films. a) Neat and b) blend ΔA data. c) Normalized photo-induced bleaching kinetics at 515 nm. d) Schematic cascaded energy transfer from high- to low-bandgap phases.

Next, the operational stability of the neat $PEA_2Cs_4Pb_5Br_{16}$ (Figure **5a**) and $PEA_2Cs_4Pb_5Br_{16}$:$C_8$-BTBT blend (Figure **5b**) PeLEDs based on the three HILs was assessed using time-dependent luminance measurements at a constant voltage with an initial luminance ($L_0$) set at 100 cd m$^{-2}$. The devices were unencapsulated and operated inside a nitrogen glovebox. Overall, neat PeLEDs with SAM HILs are more stable than reference devices based on PEDOT:PSS with the latter undergoing catastrophic failure, which results in a short operational half-life ($T_{50}$) of ≈5 min (Figure **5a**). Neat PeLED utilizing 2PACz and Br-2PACz HILs exhibit more stable operation with the initial plateau in $L_0$ (<3 min) followed by rapid degradation resulting in $T_{50}$ of 9.5 and 11.5 min, respectively (Figure **5a**). Blend PeLEDs featuring a Br-2PACz HIL appear to be the most stable, with $L_0$ exhibiting an initial increase (<1.5 min) and then a gradual drop, yielding a $T_{50}$ ≈ 18.5 min (Figure **5b**). Blend PeLEDs based on PEDOT:PSS and 2PACz are significantly less stable and show a similar $T_{50}$ ≈ 5 min as seen for neat PeLEDs with PEDOT:PSS. The larger crystalline domains and fewer boundaries facilitated by the $PEA_2Cs_4Pb_5Br_{16}$:$C_8$-BTBT blend, together with the improved charge carrier injection facilitated by Br-2PACz HIL are believed to be the primary reasons behind the drastic improvement seen in the stability of the corresponding PeLEDs. Further studies of the degradation mechanisms are required in order to identify suitable mitigation strategies to increase lifetime and efficiency. Improved SAM HILs could also be explored to improve carrier injection and recombination further. The use of mixtures of different SAM HILs, for example, could be explored using the existing library of SAMs to fine-tune the energy alignment at the injecting interface as well as engineer its surface energetics. Such straightforward developments should enable improved control over the wetting and crystallization of the perovskite layer during processing. In addition, the developments in the deposition method used to grow the active layer can also be optimized using more sophisticated methods, such as blade-coating. Lastly, optimizing the light extraction from the PeLED through appropriate light-management techniques[54] offers further opportunities to enhance the EQE.

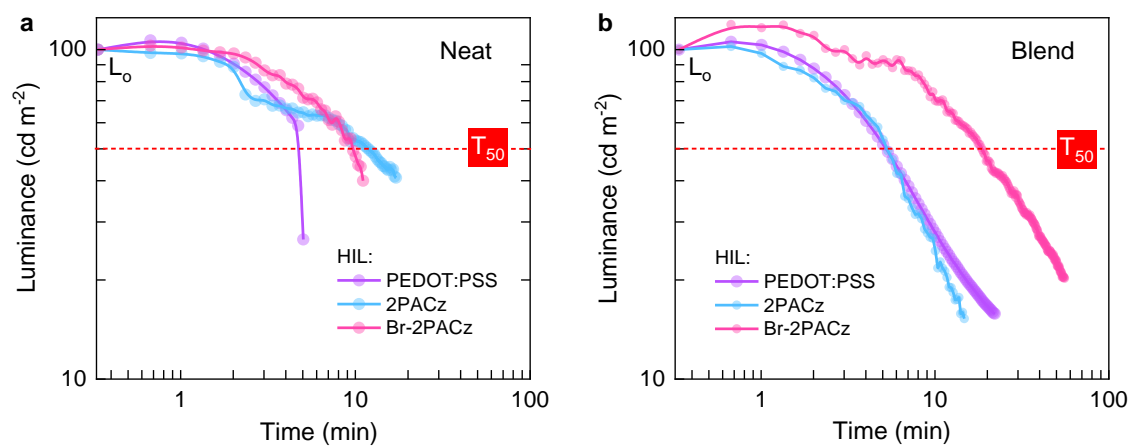

**Figure 5.** Luminance stability as a function of time for a) neat and b) blend-based PeLEDs, using PEDOT:PSS, 2PACz and Br-2PACz as hole-injection layers. The initial luminance ($L_o$) for each device is set at 100 cd/m² and its evolution is monitored as a function of time under constant biasing conditions. The operational half-life ($T_{50}$) is defined as the time taken for the luminance to drop to half of its initial value.

## 3. Summary

We developed efficient and stable green light-emitting diodes using blends of the q2D perovskite $PEA_2Cs_4Pb_5Br_{16}$ and small-molecule $C_8$-BTBT as the light-emitting and additive components, respectively. The addition of $C_8$-BTBT was found to enable control over the crystallization and growth dynamics of the $PEA_2Cs_4Pb_5Br_{16}$ phase leading to the formation of large single-crystal-like domains with a strong out-of-plane orientation. Light-emitting diodes made with the $PEA_2Cs_4Pb_5Br_{16}$:$C_8$-BTBT blends using PEDOT:PSS as the HIL exhibited higher performance than PeLEDs made with neat $PEA_2Cs_4Pb_5Br_{16}$ in terms of current efficiency (7.7 versus 4.5 cd A$^{-1}$) and EQE (3.3 versus 2.5%). Further improvements in device characteristics were achieved by replacing the PEDOT:PSS with two self-assembling monolayer HILs, namely 2PACz and Br-2PACz. SAM-based PEA$_2$Cs$_4$Pb$_5$Br$_{16}$:C$_8$-BTBT blend PeLEDs exhibited the highest current efficiency of 46.3 cd A$^{-1}$ and an EQE of 18.6% with a maximum luminance of 45 276 cd m$^{-2}$. Importantly, SAM-based PeLEDs were more stable with a longer operating lifetime. Our work demonstrates an effective method to grow high-quality q2D perovskite layers while simultaneously highlighting the tremendous potential of SAMs as carrier-injecting materials for PeLEDs.

## 4. Experimental Section/Methods

*Materials*

PEABr (99.9%), CsBr (99.9%), and PbBr$_2$ (99.9%) are procured from Sigma-Aldrich. PEDOT:PSS (AI 4083, Clevios) was purchased from Ossila and TPBi was purchased from Luminescence Technology Corp. (Lumtec). The SAMs, Br-2PACz and 2PACz, were purchased from Tokyo Chemical Industry (TCI) Co. LTD. All the above chemicals were used directly without further purification.

*Perovskite Precursor*

The 0.5 m PEA$_2$Cs$_4$Pb$_5$Br$_{16}$ precursor was prepared by dissolving PEABr, CsBr, PbBr$_2$ (2:3:4 molar ratio) in DMSO solvent. The solution was stirred for 12 h at 65 °C in a glove box with a nitrogen environment. The C$_8$-BTBT solution of 0.02 m was prepared separately using toluene as the solvent. The blend precursor formulations were prepared by mixing perovskite/organic volume ratios (*v/v*) of 75/25.

*Device Fabrication*

Indium tin oxide (ITO, 10 Ω sq$^{-1}$, purchased from Xinyan technology Ltd.) substrates were cleaned using dilute Extran 300 detergent solution, deionized water, acetone, and isopropanol, respectively, in an ultrasonic bath for 10 min each. The cleaned substrates were UV–Ozone treated for 15 min. PEDOT:PSS as the hole injection layer was prepared by spin-coating at 4000 rpm for 30 s, followed by thermal annealing at 150 °C for 15 min. The SAM-based hole injection layers, Br-2PACz (0.5 mg mL$^{-1}$ in ethanol) and 2PAcz (0.5 mg mL$^{-1}$ in ethanol) were spin-coated in the air at a speed of 3000 for 30 s and followed by annealing at 100 °C for 10 min in air. Then the substrates were transferred into a glovebox. The perovskite emissive layer coated atop the PEDOT:PSS and/or SAMs with 700 rpm for 10 s followed by 2000 rpm for 45

s. An antisolvent, toluene, dripped onto the spinning substrate ≈15 s before spinning stopped. Samples were annealed at 70 °C for 30 min. After that, the samples were transferred into a thermal evaporation system where TPBi (60 nm), Ca (10 nm), and Al (100 nm) was deposited through a shadow mask with an active device area of 10 mm$^2$. All depositions were carried out by vacuum thermal evaporation in a system made by Angstrom Engineering at a pressure below $2 \times 10^{-6}$ Torr. The final device structure consisted of ITO/HTL/perovskite (60 nm)/TPBi (60 nm)/Ca(10 nm)/Al (100) nm.

*PeLEDs Characterization*

Fabricated devices were characterized by current–voltage–luminescence (*J*–*V*–*L*) using Keysight B2912A source meter and Konica Minolta LS-150 Luminance meter controlled with a home-made setup built-in MATLAB. EL spectra of the PeLEDs were recorded using a spectrometer (Ocean Optics, QE65000) and a 2″ integrating sphere (Thorlabs, IS236A). All the measurements were performed under N$_2$ atmosphere in the glovebox.

*GIWAXS Measurements*

GIWAXS measurements were performed at the BL11 NCD-SWEET beamline at ALBA Synchrotron Radiation Facility (Spain). The incident X-ray beam energy was 12.4 eV, while the incidence angle (αi) was set between 0.1°–0.12°. The scattering patterns were recorded using a Rayonix LX255-HS area detector, which consisted of a pixel array of 1920 × 5760 pixels (*H* × *V*) with a pixel size of 44 × 44 µm$^2$. Data were expressed as a function of the scattering vector (*q*), which was calibrated using Cr$_2$O$_3$ as a standard sample at a sample to detector distance of 200.5 mm. 2D GIWAXS patterns were corrected as a function of the components of the scattering vector with a Matlab script designed for this purpose (Aurora Nogales, Edgar Gutiérrez, (2019), 2D representation of a wide angle X-ray scattering pattern as a function of *Q* vector components).

*Atomic Force Microscopy*

The surface topography information for all the samples was measured via tapping mode AFM using a Bruker dimension-icon system in the air at room temperature. The measurements were performed using Pt/Ir coated Si tips of resonance frequency ranging between 150–300 Hz.

*UV–Vis and PL Spectroscopy*

Absorption spectra were obtained using an Agilent Cary 5000 UV–vis–NIR spectrometer. Steady-state PL measurements were carried out using ambient conditions of a Horiba Aramis Raman system ($\lambda$ = 325–1064 nm, $P \leq$ 300 mW, LASER).

*Hyperspectral Imaging*

Hyperspectral PL images were recorded with an IMA hyper-spectral microscope from Photon ETC. A 405 nm CW laser was used as an excitation source and spectral images were collected from 405 to 450 nm on an area of 330 µm × 442 µm with 10 s integration time for 1 nm steps.

*TA Spectroscopy*

The fs-TA spectroscopy was performed using a multipass amplified Ti:sapphire laser (Astrella from Coherent; which generates 800 nm modelocked laser pulses with ≈100 fs pulse width having 7 mJ per pulse energy at a 1 kHz repetition rate) and in conjunction with Helios spectrometers (Ultrafast system) in a time frame of 0.1 ps to 6 ns. A fraction of the amplified pulses were directed toward a spectrally tunable (240−2600 nm) optical parametric amplifier (TOPAS, Light Conversion) to generate the excitation pump pulses (350 and 490 nm), and the pump fluences were controlled (0.8–8 µJ cm$^{-2}$) using a neutral density (ND) filter. Other fractions of the amplified 800 nm seed pulses were passed through a motorized delay stage and focused on 2-mm thick calcium fluoride (CaF$_2$) crystal to generate white light continuum (350–910 nm), which was used as probe. The pump pulses were spatially overlapped with probe pulses on the sample surface after being directed into a synchronized mechanical chopper (500 Hz). In order to improve the signal-to-noise ratio the probe pulses were split into two beams, which are commonly known as signal and reference. Change in absorption ($\Delta A = A_{\text{pump on}} - A_{\text{pump off}}$) was detected as a function of time and energy. All spectra were averaged over a time of 2 s for each time delay. The transient state was detected based on the movement of delay stage and the experimental data was analyzed using Surface Xplorer software.

*XPS*

XPS measurements were performed with a Thermofisher Escalab 250Xi system. All samples were measured under ultrahigh vacuum ($10^{-10}$ mbar). XPS measurements were performed using an XR6 monochromated AlK$\alpha$ source ($h\nu$ = 1486.6 eV) and a pass energy of 20 eV. Argon cluster ion beam etching experiments were performed using a MAGCIS ion gun using cluster energy of 4000 eV.

*SEM*

The SEM topology images of neat and blend perovskite films on ITO/PEDOT:PSS, ITO/2PACz, and ITO/Br-2PACz surfaces were acquired by an ultra-high-resolution field emission Magellan SEM equipped with a 2-mode final lens (immersion and field-free). The high-resolution images were obtained using the immersion mode at 5–10 kV after appropriate beam and lens alignments.

**Supporting Information**

Supporting Information is available from the Wiley Online Library or from the author.

Acknowledgements

M.G. and D.G. contributed equally to this work. This publication is based in part on work supported by the King Abdullah University of Science and Technology (KAUST) Office of Research Administration (ORA) under Award No: OSR-2016-CRG5-3029 and OSR-CCF-3079. Baseline funding from KAUST is also acknowledged by DDCB and TDA. This project has received funding from the European Research Council (ERC) under the European Union's Horizon 2020 research and innovation programme (ERC Grant Agreement n° 714067, ENERGYMAPS) and the Deutsche Forschungsgemeinschaft (DFG) in the framework of the Special Priority Program (SPP 2196) project PERFECT PVs (#424216076).